\documentclass[12pt]{article}
\begin{document}
\def\eps{\epsilon}
\def\epsnn{\epsilon_{NN}}
\def\la{\Lambda}
\def\si{\Sigma}
\def\sim{{\Sigma^-}}

\noindent
\centerline{\bf Maximum mass of neutron stars with a quark core}
\vskip 1 cm
\centerline{\it G. F. Burgio$^{1}$, M. Baldo$^{1}$, P. K. Sahu$^{1}$,
A. B. Santra$^2$
and H.-J. Schulze$^3$}
\vskip 0.3 cm
\centerline{$^{1}$Istituto Nazionale di Fisica Nucleare, Sezione di Catania}
\centerline{Corso Italia 57, I-95129 Catania, Italy}
\vskip 0.3 cm
\centerline{$^2$Nuclear Physics Division, Bhabha Atomic Research Center,
Mumbai 400 085, India}
\vskip 0.3 cm
\centerline{$^3$Departament d'Estructura i Constituents de la Mat\`eria,}
\centerline{ Universitat de Barcelona,
Av. Diagonal 647, E-08028 Barcelona, Spain}

\newpage
\vskip 3 cm
\noindent
{\bf Abstract.}
Massive  neutron  stars (NS) are expected to possess a quark core. While
the hadronic side of the NS equation of state (EOS)  can  be  considered
well established, the quark side is quite uncertain. While calculating the
EOS  of  hadronic  matter  we  have  used  the Brueckner-Bethe-Goldstone
formalism with realistic  two-body  and  three-body  forces, as well as
a relativistic mean field model.  For  quark
matter  we  employ  the  MIT  bag model constraining the bag constant by
exploiting the recent experimental  results  obtained  at  CERN  on  the
formation of a quark-gluon plasma. We calculate the structure of
NS  interiors  with  the EOS comprising both phases, and we find 
that the NS maximum masses fall in a  relatively  narrow  interval,  
$1.45\,M_\odot  \leq M_{\rm  max}  \leq  1.65\,M_\odot$,  near  the  lower  
limit  of  the observational range.


\newpage

An ongoing active research area, both theoretical and experimental, concerns
the  properties  of  matter  under  extreme  conditions  of  density and
temperature, and the determination of the EOS associated  with  it.  Its
knowledge  is  of  key  importance  for building models of neutron stars
(NS's) \cite{shapiro}. The observed NS  masses  are  typically  $\approx
(1-2)  M_\odot$  (where  $M_\odot$  is  the  mass of the sun, $M_\odot =
1.99\times 10^{33}$g), and the radius is of the  order  of  10  km.  The
matter in the core possesses densities ranging from a few times $\rho_0$
$(\approx  0.17\;{\rm  fm}^{-3}$,  the normal nuclear matter density) to
one order of magnitude higher. Therefore, a detailed  knowledge  of  the
EOS  is required for densities $\rho \gg \rho_0$, where a description of
matter only in terms of nucleons and leptons may be inadequate. In fact,
at densities $\rho \gg \rho_0$ several species of other particles,  such
as  hyperons  and  $\Delta$ isobars, may appear, and meson condensations
may take place; also, ultimately, at very high densities, nuclear matter
is expected to undergo a transition to a quark-gluon plasma \cite{quark}.
However, the exact value of the transition density to  quark  matter  is
unknown and still a matter of recent debate.

In this letter, we propose to constrain the maximum mass of neutron stars
taking  into  account the phase transition from hadronic matter to quark
matter inside the neutron star. For this purpose, we describe the hadron
phase   of   matter   by using two different equations of state, {\it i.e.}
a   microscopic   EOS   obtained   in   the
Brueckner~-Bethe~-Goldstone   (BBG)   theory   \cite{book},  
and  a more phenomenological relativistic mean field model \cite{serot}. 
The deconfined quark phase is treated within the popular MIT bag model
\cite{chodos}.
The bag constant, $B$, which  is  a  parameter  of  the  bag  model,  is
constrained  to  be  compatible  with  the  recent  experimental results
obtained at CERN on the formation of a quark-gluon plasma \cite{heinz},
recently confirmed by RHIC preliminary results \cite{blaizot}.
This statement requires some clarification.
In general, it is not obvious if the informations on 
the nuclear EOS from high energy heavy ion
collisions can be related to the physics of neutron stars interior. The 
possible quark-gluon plasma produced in heavy ion collision is expected to
be characterized by small baryon density and high temperature, while the 
possible quark phase in neutron stars appears at high baryon density and
low temperature. However, if one adopts for the hadronic phase
a non-interacting gas model of nucleons, antinucleons and
pions, the original MIT bag model  predicts that the 
deconfined phase occurs at an almost constant value of the
quark-gluon energy density, {irrespective} of the thermodynamical
conditions of the system \cite{gavai}. For this reason, it is
popular to draw the transition line between the hadronic and quark phase 
at a constant value of the energy density, which was estimated to 
fall in the interval between 0.5 and 2 GeV fm$^{-3}$ \cite{mul}. 
This is consistent with the value of about 1 GeV fm$^{-3}$ reported 
by CERN experiments.  \par
In this exploratory work
we will assume that this is still valid, at least approximately,
when correlations in the hadron
phase are present. We will then study
the predictions that one can draw from this hypothesis on neutron star 
structure. Any observational data on neutron stars in disagreement 
with these predictions 
would give an indication on the accuracy of this assumption.
Indeed, the hadron phase EOS can be considered well established.
The main uncertainty is contained in the quark phase EOS, 
since it can be currently described only by phenomenological models
which contain few adjustable parameters. In the case  
of the MIT bag model, which is adopted in this work, the parameters 
are fixed to be compatible with the CERN data, according to the 
hypothesis of a constant
energy density along the transition line. In practice, this means
that all our calculations can be limited to zero temperature.
\par
We  start  with the description of the hadronic phase. It has been shown
that the non-relativistic BBG expansion is well convergent  \cite{song},
and  the Brueckner-Hartree-Fock (BHF) level of approximation is accurate
in the density range relevant for neutron  stars.  In  the  calculations
reported  here  we  have  used  the  Paris potential \cite{lac80} as the
two-nucleon  interaction  and  the  Urbana  model  as  three-body  force
\cite{schi}.  This  allows  the  correct  reproduction  of the empirical
nuclear matter saturation point $\rho_0$ \cite{bbb}. Recently the  above
procedure  has  been  extended  to the case of asymmetric nuclear matter
including  hyperons   \cite{bbs,barc}   by   utilizing   hyperon-nucleon
potentials that are fitted to the existing scattering data.
\par
To complete our analysis, we will consider also a hadronic EOS
derived from relativistic mean field model (RMF) \cite{sahu}. 
The BHF and the RMF EOS are both shown in Fig.1.
The parameters of the RMF model have been taken in such a way that the
compressibility at saturation is around $260$ MeV, 
the same as in BHF calculations and close to estimates from monopole
oscillations in nuclei \cite{tole}. 
The symmetry energy is also quite similar for the two EOS, about
30 MeV at saturation.
\par
For the deconfined quark phase,  within the MIT bag model
\cite{chodos}, the total energy density is the sum of a non-perturbative
energy shift $B$, the bag constant, and the kinetic energy
for non-interacting massive quarks of
flavors $f$ with mass $m_f$  and  Fermi  momentum  $k_F^{(f)}  [=  (\pi^2
\rho_f)^{1/3}$, with  $\rho_f$ as the quarks' density of flavour $f$] 
\begin{equation}
{E  \over V} = B + \sum_f {3m_f^4 \over 8\pi^2} \Big[ x_f \sqrt{x_f^2+1}
\left(2x_f^2 + 1\right) - {\rm sinh^{-1}}x_f \Big] \:,
\label{e:eos}
\end{equation}
where  $x_f  = k_F^{(f)}/m_f$. 
We consider in this work massless $u$ and
$d$ quarks, whereas the $s$ quark  mass  is  taken  equal  to  150  MeV.
The bag constant $B$ can be interpreted as the difference
between the energy densities of the perturbative vacuum and the physical
vacuum.
Inclusion of perturbative interaction among quarks introduces additional
terms  in  the  thermodynamic  potential \cite{fahri} and hence in the
number density and the energy density; however, when taken into  account
in  the  first order of the strong coupling constant, these terms do not
change our results appreciably. Therefore, in order to calculate the EOS
for quark matter we restrict to Eq.~(\ref{e:eos}). In the  original  MIT
bag   model   the   bag  constant  has  the  value  $B  \approx  55\,\rm
MeV\,fm^{-3}$,  which  is  quite  small  when  compared  with  the  ones
($\approx  210\,\rm  MeV\,fm^{-3}$)  estimated from lattice calculations
\cite{grei}. In this sense $B$ can be considered as a free parameter.

We try to determine a range of possible values for $B$ by exploiting the
experimental  data  obtained  at the CERN SPS, where several experiments
using high-energy beams of Pb nuclei reported (indirect) evidence  for
the  formation  of  a  quark-gluon  plasma  \cite{heinz}.  The resulting
picture is the following: during  the  early  stages  of  the  heavy-ion
collision,  a very hot and dense state (fireball) is formed whose energy
materializes in the form of quarks and gluons strongly interacting  with
each  other,  exhibiting  features  consistent  with expectations from a
plasma of deconfined quarks  and  gluons \cite{satz}.  Subsequently,  the
``plasma"  cools  down and becomes more dilute up to the point where, at
an energy density of about $1\,\rm GeV\, fm^{-3}$ and  temperature  
$T\approx
170\,\rm  MeV$,  the  quarks and gluons hadronize. The expansion is fast
enough so that no mixed hadron-quark equilibrium phase  is  expected  to
occur, and no weak process can play a role. According to the analysis of
those  experiments,  the  quark-hadron  transition  takes place at about
seven times  normal  nuclear  matter  energy  density  ($\eps_0  \approx
156\,\rm MeV\, fm^{-3}$).
In  the  MIT  bag  model,  the structure of the QCD phase diagram in the
chemical potential and temperature  plane  is  determined  by  only  one
parameter,  $B$,  although  the  phase  diagram  for the transition from
nuclear matter to quark matter  is  schematic  and  not  yet  completely
understood, particularly in the light  of  recent  investigations  on  a
color  superconducting  phase  of quark matter \cite{khrisna}. 
As discussed above, in our analysis we assume that the transition to 
quark-gluon plasma 
is determined by the value of the energy density only (for a given asymmetry). 
With  this assumption  and  taking  the  hadron  to  quark matter 
transition energy density from the CERN experiments we estimate the value 
of $B$ and its possible density dependence as given below.
\par
First,   we  calculate  the  EOS  for  cold  asymmetric  nuclear  matter
characterized by a proton fraction $x_p = 0.4$ (the one  for  Pb  nuclei
accelerated at CERN-SPS energies) in the BHF formalism with two-body and
three-body forces as described earlier. The result is shown by the solid
line  in  Fig.~1a). Then we calculate the EOS for $u$ and $d$
quark matter using Eq.~(\ref{e:eos}). We find that at  very  low  baryon
density  the  quark matter energy density is higher than that of nuclear
matter, while with increasing baryon density the  two  energy  densities
become  equal at a certain point [indicated in Fig.~1a) by the full dot)],
and
after that the nuclear matter energy density remains always  higher.  We
identify  this  crossing  point with the transition density from nuclear
matter  to  quark  matter.  
To be more precise, this crossing fixes the density interval where the phase 
transition takes place. In fact, according to the Gibb's construction,
the crossing must be located at the center of the mixed phase region,
if it is present.
To  be  compatible  with  the   experimental
observation  at  the  CERN-SPS,  we  require  that  this  crossing point
corresponds to  an  energy  density  of  $E/V  \approx  7\eps_0  \approx
1.1\,\rm  GeV\,fm^{-3}$.  However, for no density independent
value of $B$, the two  EOS'  cross  each  other,  satisfying  the  above
condition.  Therefore, we try a density dependent $B$. In the literature
there  are  attempts  to  understand  the  density  dependence  of   $B$
\cite{brown};  however, currently the results are highly model dependent
and no definite picture has come  out  yet.  Therefore,  we  attempt  to
provide  effective  parametrizations for this density dependence, trying
to  cover  a  wide  range  by  considering  some  extreme  choices.  Our
parametrizations  are  constructed  in  such  a  way  that at asymptotic
densities $B$ has some finite value $B_{as}$. 
  We  have  found  $B_{as}  =  50\,\rm
MeV\,fm^{-3}$ for the BHF case, but have verified that our results do not 
change appreciably by varying this value, since at large densities the quark
matter   EOS   is   dominated   by  the  kinetic  term  on  the  RHS  of
Eq.~(\ref{e:eos}). First, we use a Gaussian parametrization given as
\begin{eqnarray}
B(\rho)  =  B_{as}  +  (B_0  -  B_{as})  \exp  \left[  -\beta  \Big(
\frac{\rho}{\rho_0} \Big)^2 \right] \:. \label{e:g}
\end{eqnarray}
The parameter $\beta$ has been fixed by equating the quark matter energy
density  from  Eq.~(\ref{e:eos})  with  the nucleonic one at the desired
transition density $\rho_c = 0.98\,{\rm fm^{-3}}$ (represented by the 
full dot in Fig. 1a)), i.e.  $E/V(\rho_c) =  1.1\,\rm  GeV\,fm^{-3}$.
Therefore  $\beta$  will  depend  only  on  the  free  parameter  $B_0 =
B(\rho=0)$. However, the exact value of $B_0$ is not very  relevant  for
our  purpose,  since  at  low  density  the matter is in any case in the
nucleonic phase. We attempt to cover the  typical  range  by  using  the
values  $B_0  =  200\,\rm  MeV\,fm^{-3}$ and $400\,\rm MeV\,fm^{-3}$, as
shown in Fig.~1c). We also use another  extreme,  Woods-Saxon
like, parametrization,
\begin{eqnarray}
B(\rho)  =  B_{as} + (B_0 - B_{as}) \left[1 + \exp\left(\frac{\rho -
\bar\rho}{\rho_d} \right)\right]^{-1} \:, \label{e:ws}
\end{eqnarray}
where $B_0$ and $B_{as}$ have the same meaning as described before for
Eq.~(\ref{e:g}) and $\bar\rho$ has been fixed in the same way as $\beta$
for  the previous parametrization. For $B_0 = 400\,\rm MeV\,fm^{-3}$, we
get  $\bar\rho  =  0.8\,\rm  fm^{-3}$ for $\rho_d  =  0.03\,\rm  fm^{-3}$.
With  this parametrization $B$
remains practically constant at a value $B_0$ up to  a  certain  density
and  then  drops  to $B_{as}$ almost like a step function, as shown by
the  long-dashed  curve  in  Fig.~1c).  It  is   an   extreme
parametrization  in  the sense that it will delay the onset of the quark
phase in neutron star matter as much as possible. Both  parametrizations
Eqs.~(\ref{e:g})  and  (\ref{e:ws})  yield  the  transition from nuclear
matter to quark  matter  at  the  energy  density  compatible  with  the
experiments.
\par
The same procedure has been followed for the RMF EOS,
see Figs. 1b) and 1d). In this case
the parameter $B_{as} $ is slightly smaller, about $38$ MeV fm$^{-3}$.
With  these  parametrizations  of  the  density dependence of $B$ we now
consider  the  hadron-quark  phase  transition  in  neutron  stars.   We
calculate in the BHF framework and in the RMF approach the EOS of a 
conventional neutron star as
composed  of  a  chemically  equilibrated  and charge neutral mixture of
nucleons, leptons and hyperons. The  result  is  shown  by  the  
solid  lines  in Fig.~2a) and 2b) respectively.
The other curves (with the same notation
as in Fig.~1) represent the EOS' for  beta-stable  and  charge
neutral  quark  matter. 
We determine the range of baryon density where both
phases can coexist by following the construction from ref. \cite{glen}.
In this procedure both hadron and quark phases are allowed to be charged,
still preserving the total charge neutrality. Pressure is the same in the two 
phases to ensure mechanical stability, while the chemical potentials of
the different species are related to each other to ensure chemical and
beta stability. The resulting EOS for neutron star matter, according to
the different bag parametrizations, is reported in Fig. 3, where the shaded
area indicates the mixed phase region. 
A pure quark phase is present at  densities
above the shaded area  and a pure hadronic phase is present below it.
The onset density of the mixed phase turns out to be slightly
smaller than the density for hyperons formation in pure hadronic
matter. Of course hyperons are still present in the hadron component
of the mixed phase. For the Wood-Saxon parametrization of the
bag constant the mixed phase persists up to high baryon density.
As previously anticipated, this is in agreement with the delayed crossing 
of the energy density curves for the hadron and quark phases, as can be 
seen from Fig. 2. 
Finally,    we    solve    the    Tolman-Oppenheimer-Volkoff   equations
\cite{shapiro}  for  the  mass  of  neutron  stars  with  the  EOS'   of
Figs.~3  as  input.  The  calculated  results,  the NS mass vs.
central density, for all cases are shown in Fig.~4a) and 4b). The EOS with
nucleons, leptons and hyperons gives a maximum mass of neutron stars of 
about $1.26\,M_\odot$ in the BHF case.  
In the case of the RMF model, the corresponding EOS produces values of 
the maximum mass close to $1.7\,M_\odot$.    
It is commonly believed that the inclusion of the quark component
should soften the NS matter EOS. This is indeed the case in the 
RMF model, as apparent in Fig. 4b). However the situation is reversed
in the BHF case, where the  EOS  becomes, on the contrary, stiffer.
Correspondingly, the inclusion of the quark component has the
effect of increasing the maximum mass in the BHF case and of decreasing it
in the RMF case. As a consequence, 
the calculated maximum masses fall in any case in a relatively
narrow range, $1.45\,M_\odot \leq M_{\rm max} \leq 1.65\,M_\odot$,
slightly above the observational lower limit of $1.44\,M_\odot$ \cite{hulse}.
\par
As one can see from Fig. 4, the presence of a mixed phase produces
a sort of plateau in the mass vs. central density relationship,
which is a direct consequence of the smaller slope displayed by
all EOS in the mixed phase region, see Fig. 3. In this region,
however, the pressure is still increasing monotonically, 
despite the apparent smooth behaviour, and no unstable configuration can 
actually appear. We found that the appearance of this slow variation
of the pressure is due to the density dependence of the bag constant,
in particular the occurrence of the density derivative of the
bag constant in the pressure and chemical potentials, as required
by thermodynamic consistency. To illustrate this point
we calculate the EOS for quark matter with a density independent
value of $B = 90$ MeV fm$^{-3}$, see Fig. 5, and the corresponding
neutron star masses. The EOS is now quite smooth and the mass
vs. central density shows no indication of a plateau. More details
on this point will be given elsewhere \cite{fut}. 
\par
Finally, it has to pointed out that the maximum mass value, whether B is
density dependent or not, is dominated by the quark EOS at densities where
the bag constant is much smaller than the quark kinetic energy.
The constraint coming from heavy ion reactions, as discussed
above, is relevant only to the extent that it restricts 
B at high density within a range of values, which are commonly
used in the literature. This can be seen also from Fig. 5, where 
the (density independent) value of $B = 90 MeV$ produces again
a maximum value around $1.5$ solar mass.  
\par
In  conclusion,  under our hypothesis,  we found first that a density 
dependent $B$ is
necessary to understand the CERN-SPS findings on  the  phase  transition
from  hadronic  matter  to  quark matter. Then, taking into account this
observation, we calculated NS maximum masses, using  an  EOS  which
combines reliable EOS's for hadronic matter and a bag model EOS for
quark matter. The calculated maximum NS masses lie in a narrow range  in
spite of using very different parametrizations of the density dependence
of  $B$.  Other recent calculations of neutron star properties employing
various RMF nuclear EOS' together with either effective mass bag  
model  \cite{bag}  or
Nambu-Jona-Lasinio  model  \cite{njl}  EOS'  for quark matter, also give
maximum masses of only about $1.7\,M_\odot$, even though not constrained
to reproduce simultaneously the CERN-SPS data. 
The value of the maximum mass of neutron stars obtained according to
our analysis appears robust with respect to the uncertainties 
of the nuclear EOS. 
Therefore, the experimental observation of a heavy 
($M > 1.6 M_\odot$) neutron star, as claimed recently by some groups
\cite{kaaret}( $M \approx 2.2 M_\odot$ ),
if confirmed,
would suggest mainly two possibilities. Either 
serious problems are present for the current theoretical modelling 
of the high-density phase of nuclear matter, 
or the working hypothesis 
that the transition to the deconfined phase occurs approximately at the same
energy density, irrespective of the thermodynamical conditions, is
substantially wrong. In both cases, one can expect a well defined hint on the 
high density nuclear matter EOS.

This  work  was  supported  in  part  by  the  programs  ``Estancias  de
cient\'{\i}ficos y  tecn\'ologos  extranjeros  en  Espa\~na'',  SGR98-11
(Generalitat de Catalunya), and DGICYT (Spain) No.~PB98-1247.

\newpage

\begin{figure}
{\bf Figure captions.}
\caption{(a,b) The  energy  density  $E/V$  vs.~the  baryon
density  $\rho$  for  nuclear matter and quark matter of charge fraction
$x_p=0.4$. The dot indicates the common intersection of the curves.  
(c,d) Density  dependence  of  the  bag  constant $B$ (see text for details).}
\caption{The energy density vs. baryon density for pure hadron matter 
(full lines) are reported for the BHF (left panel) and 
RMF (right panel) schemes, in comparison with the quark energy densities
(broken lines) with different parametrizations of the bag constant.}
\caption{Total EOS including both hadronic and quark components. 
Different prescriptions for the quark phase are considered, see the text
and Figs. 1 and 2.
Fo the hadron component the BHF (left panel) and the RMF (right panel)
schemes are considered. In all cases the shaded region indicates the mixed 
phase MP, while HP and QP label the portion of the EOS where pure hadron and 
pure quark phases, respectively, are present.} 
\caption{The gravitational mass of neutron stars  vs.~the
central  density  for  the  EOS's  shown in Fig.3.} 
\caption{In the left panel is shown the EOS for neutron star matter 
(dashed lines labeled by HP + QP) for a density independent 
value of the bag constant $B = 90 MeV\,$fm$^{-3}$, with BHF (a) anf RMF (c) 
hadron equations of state. The shaded areas indicate the mixed phase region.
The corresponding masses vs. central density are shown on the right panels. 
In all cases the thin and thick lines correspond to the results obtained 
for pure quark and pure hadron EOS respectively. }
\end{figure}

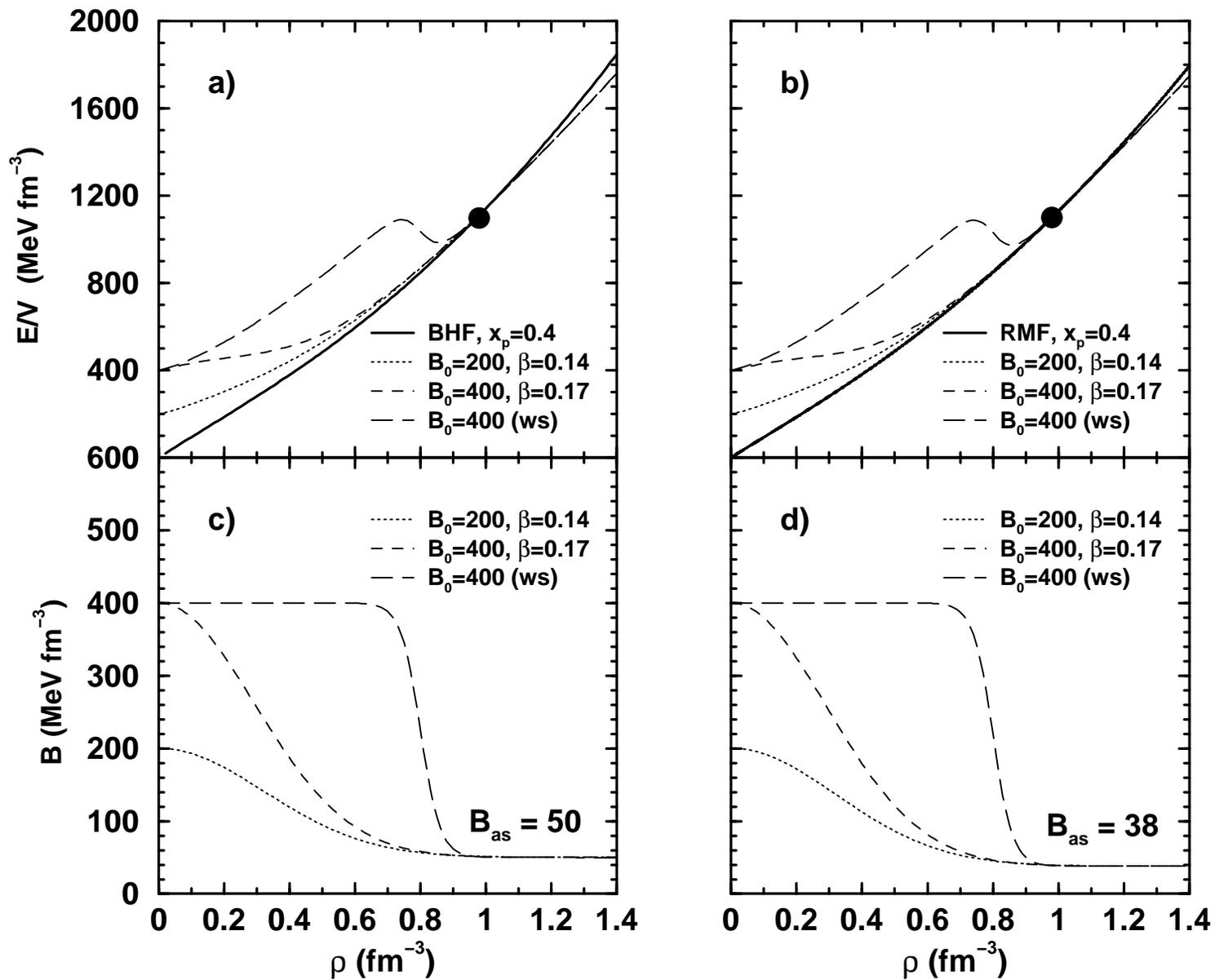
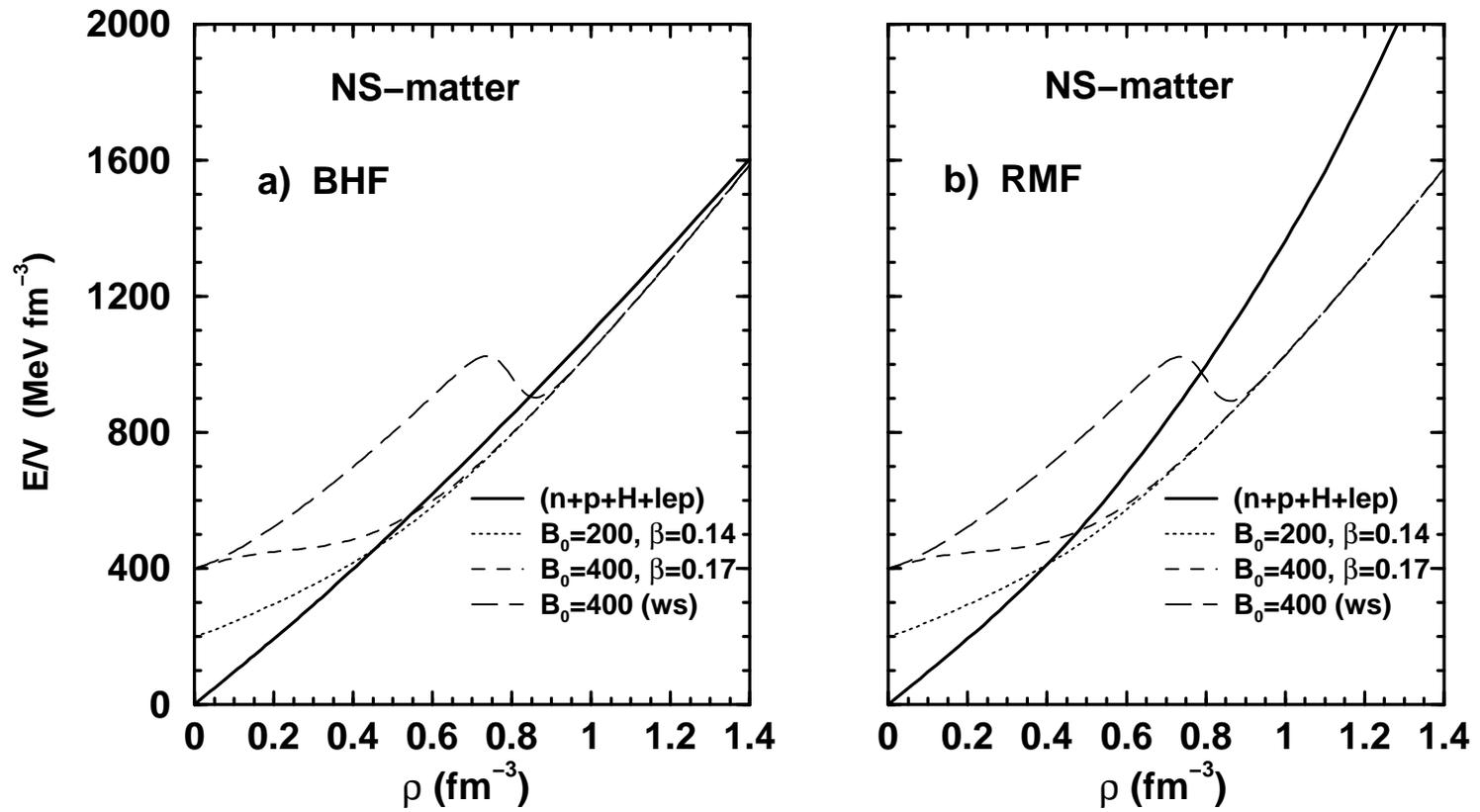
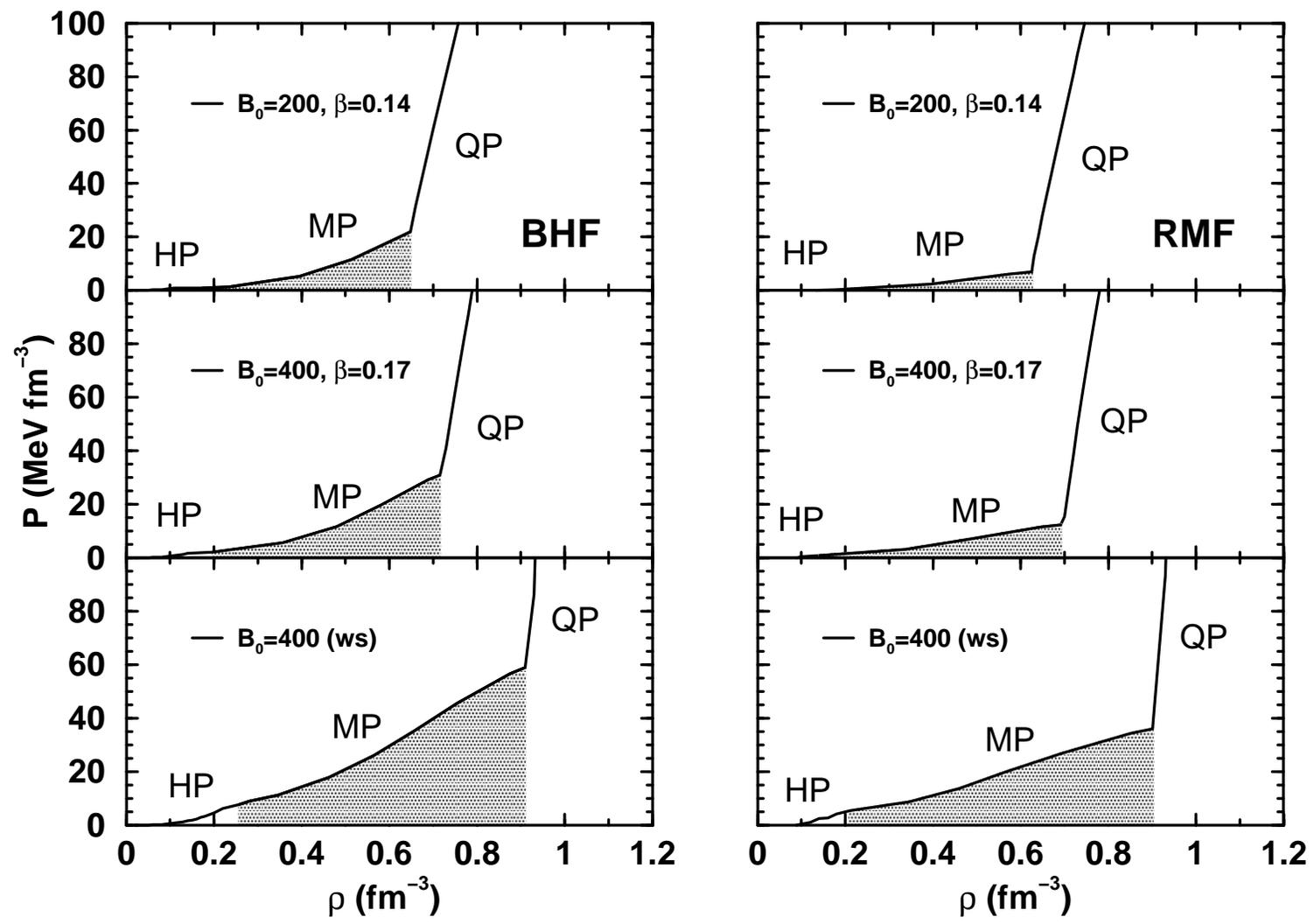
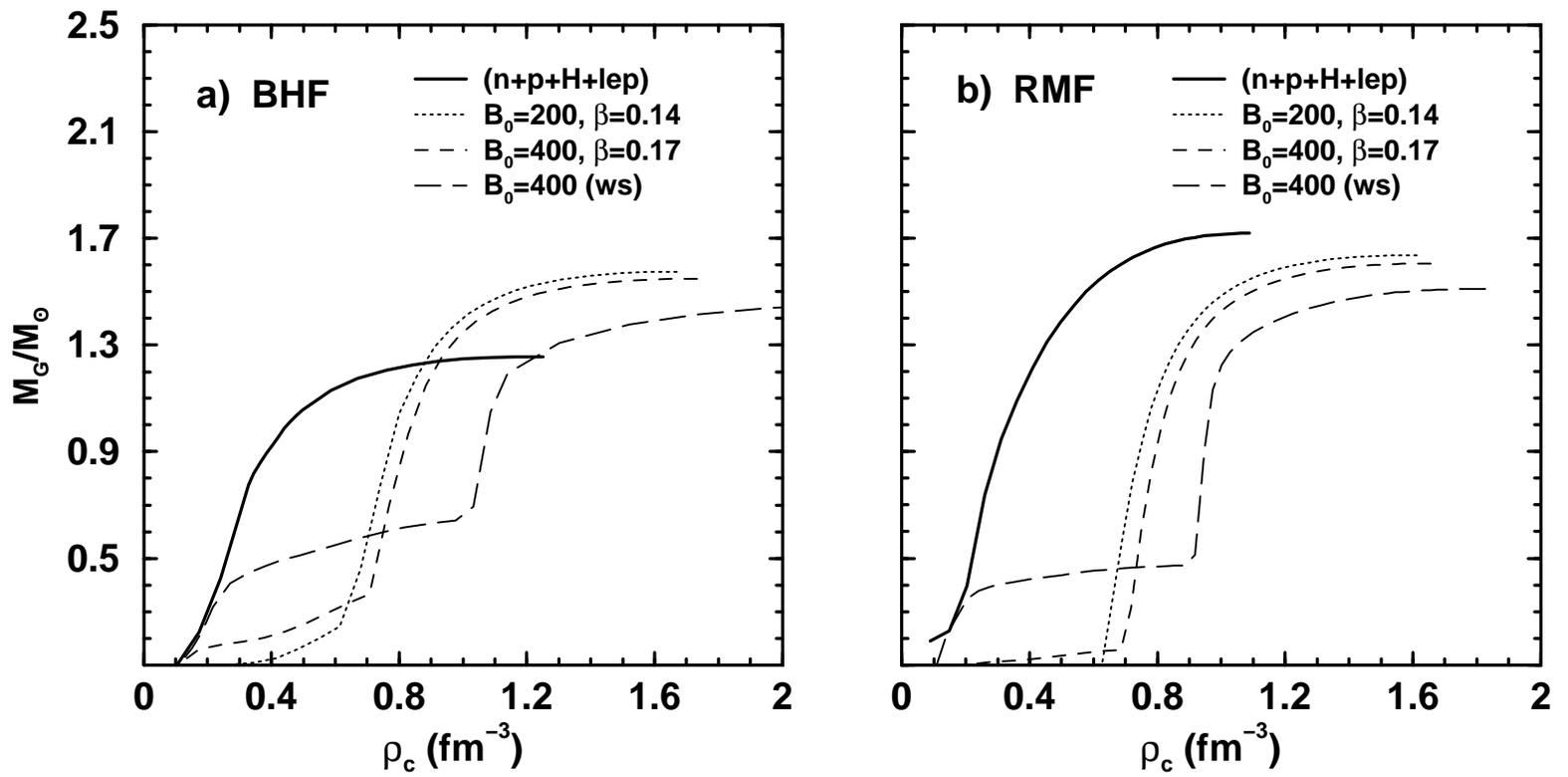
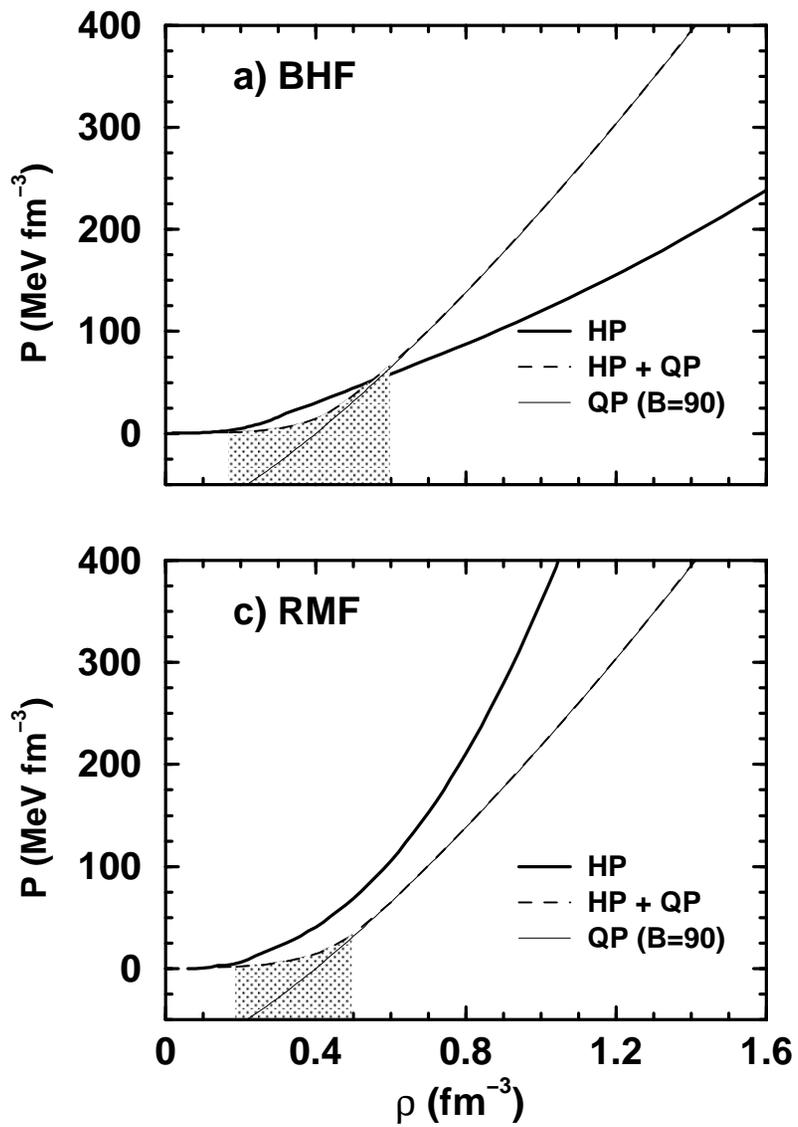
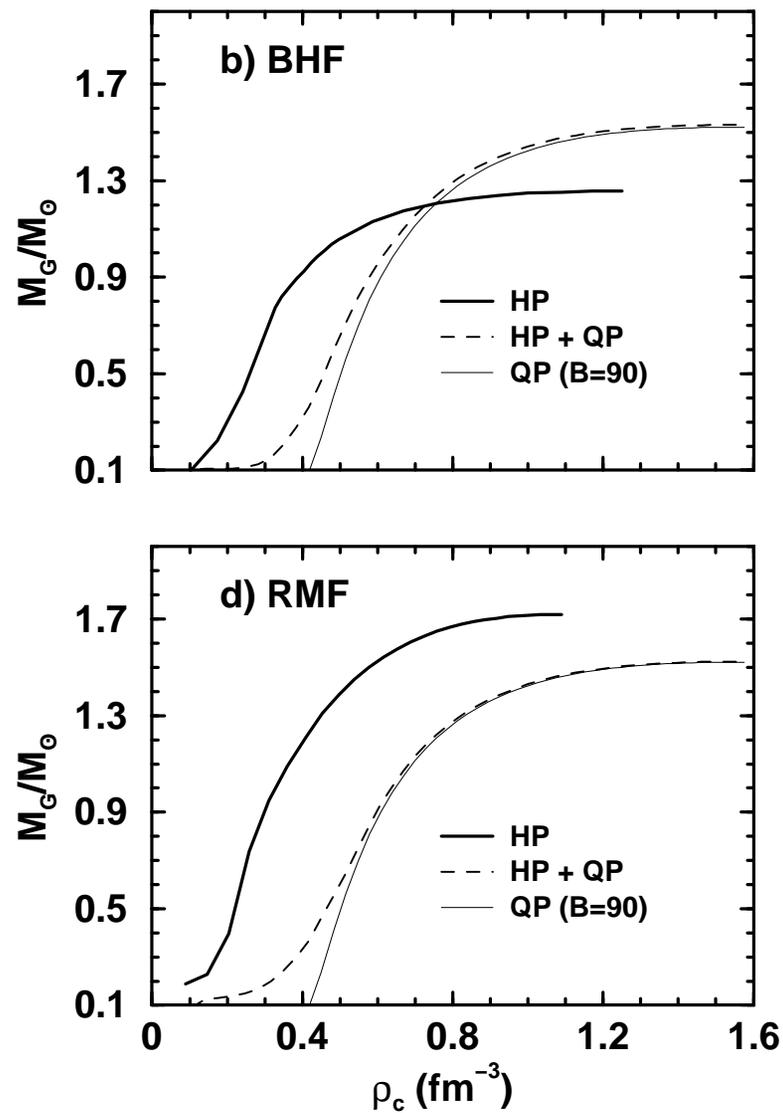
\end{document}